\documentclass[sigconf]{acmart}

\setcopyright{acmlicensed}
\copyrightyear{2026}
\acmYear{2026}
\acmDOI{}
\acmConference[FutureHCI '26]{The Future of HCI Workshop}{August 17-18, 2026}{Blacksburg, VA}
\acmISBN{}

\makeatletter
\newcommand{\@publicationmetadataspace}[3]{%
  \leavevmode\hbox to #1sp{\vrule width 0pt height #2sp depth #3sp\hfil}}
\patchcmd{\maketitle}
  {\ifx\@acmISBN\@empty\else ACM~ISBN~\@acmISBN\\\fi}
  {\ifx\@acmISBN\@empty
     \@publicationmetadataspace{7747844}{300941}{27066}\\%
   \else ACM~ISBN~\@acmISBN\\\fi}
  {}{\PackageError{paper}{Could not preserve ISBN spacing}{Check the bundled acmart class.}}
\patchcmd{\maketitle}
  {\ifx\@acmDOI\@empty\else\@formatdoi{\@acmDOI}\fi}
  {\ifx\@acmDOI\@empty
     \@publicationmetadataspace{7166150}{315162}{108265}%
   \else\@formatdoi{\@acmDOI}\fi}
  {}{\PackageError{paper}{Could not preserve copyright DOI spacing}{Check the bundled acmart class.}}
\patchcmd{\@mkbibcitation}
  {\ifx\@acmDOI\@empty\else\@formatdoi{\@acmDOI}\fi}
  {\ifx\@acmDOI\@empty
     \@publicationmetadataspace{8189899}{360186}{123732}%
   \else\@formatdoi{\@acmDOI}\fi}
  {}{\PackageError{paper}{Could not preserve reference DOI spacing}{Check the bundled acmart class.}}
\makeatother

\begin{document}

\title{Automating the Application of HCI Principles}

\subtitle{Skills for On-Demand UI Construction, the Human-AI Space to Think, and
the Future of HCI}

\author{Nathan Conklin}
\email{nathan.conklin@vt.edu}
\orcid{0009-0002-9276-6777}
\affiliation{%
  \institution{Virginia Tech}
  \city{Blacksburg}
  \state{Virginia}
  \country{USA}
}

\author{Miranda Capra}
\email{mcapra@gmail.com}
\orcid{0009-0004-6920-1174}
\affiliation{%
  \institution{Independent Consultant}
  \city{Chapel Hill}
  \state{North Carolina}
  \country{USA}
}

\author{Chris North}
\email{north@cs.vt.edu}
\orcid{0000-0002-8786-7103}
\affiliation{%
  \institution{Virginia Tech}
  \city{Blacksburg}
  \state{Virginia}
  \country{USA}
}

\begin{abstract}
Human-computer interaction (HCI) is in the middle of a transition: large
language models can now generate functional user interfaces (UIs) on demand
from natural-language task descriptions. A user explains what they are
trying to accomplish, and the system materializes a working interface to
support it. This capability already exists in systems such as Claude and
ChatGPT and continues to grow in fidelity as the underlying models
improve. The next step along this trajectory is to move from interfaces
that are merely generated to interfaces that are \emph{generated well}.
We propose a framework in which the dialogue between user and artificial intelligence (AI) becomes
a Space to Think: a shared, structured cognitive workspace in which task
decomposition produces an on-demand user interface as an extension of
the user\textquotesingle s thinking rather than as a separate artifact.
Within this paradigm, classical HCI design knowledge
(Nielsen\textquotesingle s heuristics, Norman\textquotesingle s
affordance prescriptions, Web Content Accessibility Guidelines (WCAG)
success criteria, cognitive-load
constraints, and mixed-initiative principles) is encoded as skills:
machine-readable skill.md files that the generating agent loads at
runtime as software engineering tools. Skills turn HCI design knowledge
into declarative, inspectable, version-controlled, and editable
artifacts owned by the HCI community itself so that accessibility,
learnability, and consistency become properties of a generative process
rather than properties of a finished product. We outline a research
agenda depicting a future where the HCI field transitions from today's
design/knowledge heuristic checklist towards a future where the craft
becomes machine-readable, executable, and open.
\end{abstract}



\maketitle

\section{Introduction}

HCI has been dominated by a common design workflow: a designer
anticipates a user\textquotesingle s task, creates the layout of the
corresponding user interface, and ships the result as an application.
The user\textquotesingle s job is to translate intent into the
vocabulary the designer chose. When the user\textquotesingle s task does
not match the designer\textquotesingle s anticipation, the user pays the
cost in workarounds, rework, and abandoned features. Direct manipulation \cite{bigben1983} made this within user's reach; the web
made it ubiquitous; mobile platforms made it global. HCI primarily
remains in a regime of pre-designed interaction. Three converging shifts
now make this regime contingent rather than necessary. First, large language models (LLMs) can
generate high-fidelity, executable user interface code from natural
language with growing reliability \cite{cao2025}.
Second, agent architectures based on chain-of-thought, ReAct,
Tree-of-Thoughts, and Plan-and-Solve prompting allow models to decompose
tasks into sub-goals and act through
tools \cite{chen2025}. Third, commercial systems
including Anthropic\textquotesingle s Artifacts,
OpenAI\textquotesingle s Canvas, and Vercel\textquotesingle s v0 enable
dialogue surfaces in which an interface materializes while the user is
still describing what they want \cite{anthropic2024a}. The
component pieces of an on-demand user interface paradigm are already in production;
what is missing is a conceptual frame that explains what they are doing.

HCI-empowered software engineering will enable on-demand User Interface
construction within a Human-AI cognitive workspace. The dialogue between
the user and AI will become more than a query-answer transaction; it
will become a Space to Think \cite{andrews2010}\cite{anthropic2026a} for
building shared context and common consensus through interaction. User
capabilities will grow through Human-AI conversation. We make the
following contributions: (i) We define the Space to Think as a
structured cognitive workspace and show how task decomposition naturally
lives inside it. (ii) We describe an architecture for on-demand user interface
construction as a workspace rather than a separate artifact. (iii) We
show that classical HCI design principles can be encoded as procedural
constraints on the generator, turning learnability, accessibility, and
consistency into properties of a process rather than properties of a
product. (iv) We describe a framework with an agenda for the next year
and the next quarter-century towards the evaluation, predictability, and
accessibility of dynamically generated content.

\begin{figure*}
  \centering
 \includegraphics[width=\textwidth]{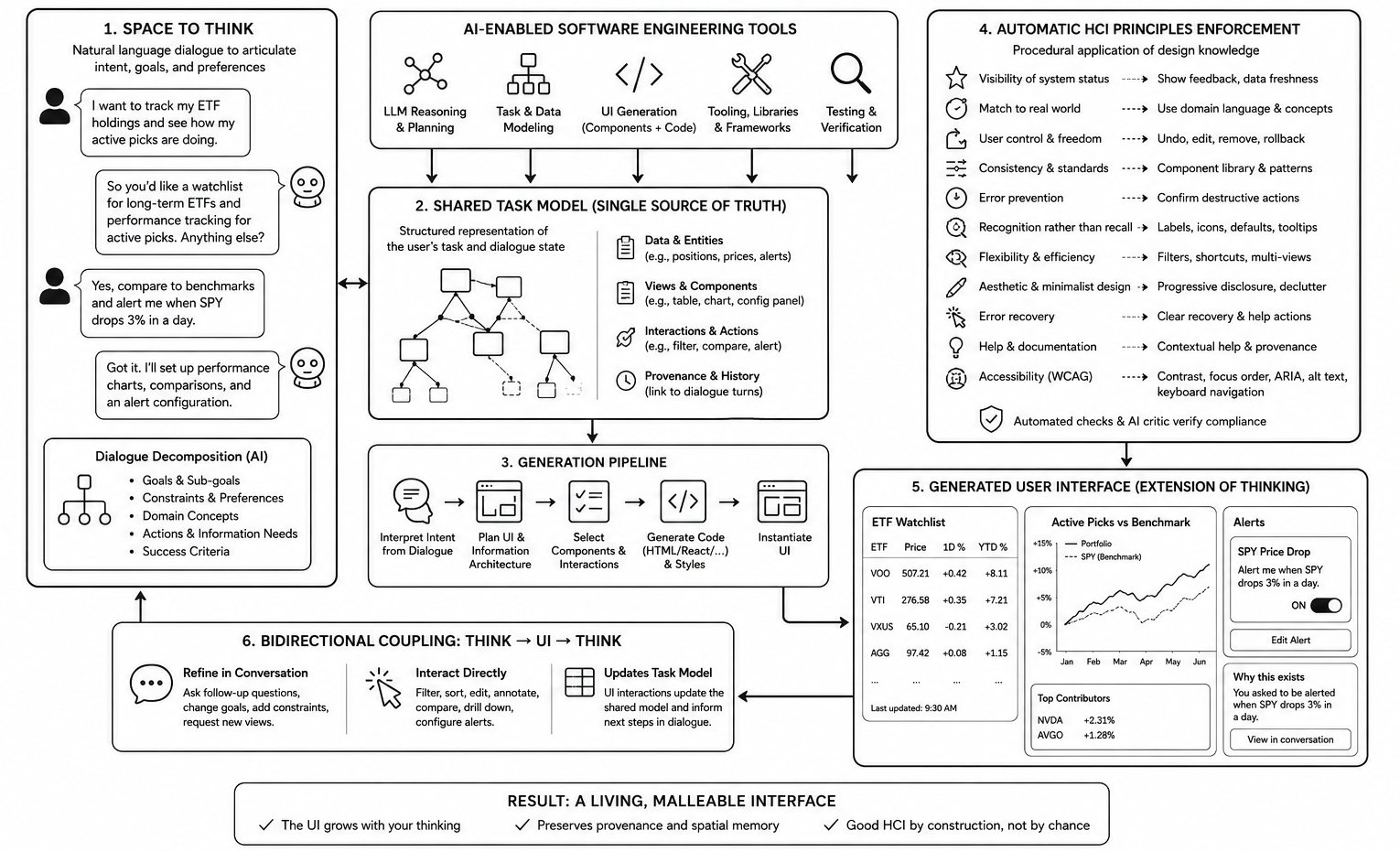}
  \caption{AI-Enabled Software Engineering That Automates UI Construction and the Application of HCI Principles}
  \Description{See section A Figure 1 Detailed Description for full text and details. The figure shows a cycle of 6 panels. (1) Space to think $\rightarrow$ (2) shared task model $\rightarrow$ to both (1) space to think and (3) generation pipeline; (3) generation pipeline $\rightarrow$ (5) generated user interface $\rightarrow$ (which is also informed by ), (6) bi-directional coupling $\rightarrow$ (1) space to think. AI-enabled software engineering $\rightarrow$ (2) shared task model. (4) automatic HCI principles enforcement $\rightarrow$ (5) generated user interface. The result is a living, malleable interface: the user interface grows with your thinking; preserves provenance and spatial memory; and has good HCI construction, not by chance.}
  \label{fig:my_diagram}
\end{figure*}

\ref{fig:my_diagram} summarizes the framework and serves as the structural backbone of the paper: (1) the Space to Think and its dialogue decomposition; (2)
the shared task model; (3) the generation pipeline; (4) automatic HCI
principles enforcement via skills; (5) the generated user interface as an extension
of thinking; (6) bidirectional coupling between dialogue and user interface.

Our primary contribution is the identification of a concrete mechanism
for procedural HCI: skills, packaged as machine-readable
\textbf{skill.md} files that the AI loads at runtime as software
engineering tools. The \textbf{skill.md} construct enables HCI design
knowledge to become a declarative, inspectable, version-controlled, and
editable software engineering tool managed by the HCI community itself.
Each skill encodes HCI design knowledge including
Nielsen\textquotesingle s heuristics \cite{nielsen1994},
Norman\textquotesingle s affordance prescriptions
\cite{norman2013a}, WCAG success criteria
\cite{w3c2024a}, cognitive-load constraints, and
mixed-initiative principles \cite{horvitz1999} as
procedural guidance that the agent can read, follow, and verify against
its own output. Where heuristic evaluation has historically been a
post-hoc audit performed by humans, skills make it an inline constraint
on  generation. Good HCI becomes a property of a process,
not a property of a product.

\section{Generative UI and LLM-Driven User Interface
Construction}

Generative UI, user interface synthesis by LLMs, has rapidly evolved.
Recent systems, such as \textsc{Jelly}, generate user interfaces around
task-driven data models that users reshape through natural language and
direct manipulation \cite{cao2025}. Generative UI
systems utilize LLMs to construct task-specific user interfaces instead
of returning linear text, enabling cognitive gains on structured and
information-dense tasks. Kyungho Lee summarizes this emerging area with
five characteristics of generative UI: computational co-creation,
expanded design-space exploration, representational fluidity, contextual
adaptation, and synthesis-over-selection
\cite{lee2025}. Parallel work including Design2Code,
UICoder \cite{wu2024}, and retrieval-augmented GUI
generation demonstrates improving fidelity in LLM-driven user interface
capabilities. Many industry capabilities have been made available on the
market. Vercel\textquotesingle s v0 operationalized text-to-React
interface generation; Anthropic\textquotesingle s Artifacts and
OpenAI\textquotesingle s Canvas introduced side-by-side conversational
workspaces in which documents, code, and user interfaces are constructed
directly from dialogue \cite{anthropic2024a}. In parallel,
the malleable software community argues that software should remain
continuously reshapeable by users rather than fixed after deployment.
Earlier model-based UI systems demonstrated that abstract task and
dialogue models could generate interfaces across heterogeneous devices;
LLM-based generation may be understood as a stochastic continuation of
this model-driven tradition.

Mixed-initiative interaction provides the conceptual foundation for
these systems. Horvitz\textquotesingle s Principles of Mixed-Initiative
User Interfaces \cite{horvitz1999} articulated how
autonomous assistance and user control should be balanced under
uncertainty, while the TRAINS project demonstrated integrated
dialogue-driven planning systems combining natural-language
understanding and task reasoning. Grosz and Sidner\textquotesingle s
discourse theory further introduced linguistic, intentional, and
attentional structures that map naturally onto collaborative cognitive
workspaces. Contemporary LLM systems inherit many of these ideas while
embedding them inside end-to-end conversational environments.

\section{On-Demand UI as Cognitive
Extension}

A Space to Think is a shared, persistent, structured workspace in which
a human user and an AI collaborator externalize, manipulate, and refine
the representations needed to solve a task
\cite{andrews2010}\cite{anthropic2026a}. The
space is shared: both participants can read and write all of its
contents, and both can hold attention on the same fragment. The space is
structured: it is a representation with internal scaffolding of
sub-goals, hypotheses, evidence, decisions, and
tools-under-construction. Finally, the space is operational for the
user. This third property is what distinguishes a Space to Think from an
ordinary chat log and what allows an on-demand UI to enable new user
interface capabilities and expand the Human-AI shared context.

Decomposition in the Space to Think provides a mixed-initiative
interaction \cite{horvitz1999}: either the human or the
AI may propose a sub-goal, contest it, or revise the plan, and
uncertainty is treated as a primary signal for when the AI should ask
rather than act. Each sub-goal that is identified via dialogue produces
a corresponding affordance in the user interface, so the structure of the
user\textquotesingle s thought is mirrored by the structure of the user
interface tool that emerges. The on-demand UI is not the output of the
Space to Think; it is an extension of it. The user interface is constructed as a
part of a shared context that enables the user to work, think, and
through use provide feedback to the AI to support future concept
iterations. Dialogue state and UI state share the same context that is
grounded in both a task-driven data model and model-based UI development
\cite{cao2025}.

\section{Automating HCI Design
Principles}

User interface construction can be automated while incorporating
critical design principles, applied procedurally to every generated
artifact to achieve a coverage that hand design rarely matches in
practice.

\textbf{Heuristics as procedural constraints:} Nielsen\textquotesingle s
ten heuristics \cite{nielsen1994} (visibility of system
status, match between system and the real world, user control and
freedom, consistency and standards, error prevention, recognition rather
than recall, flexibility and efficiency, aesthetic and minimalist
design, error recovery, and help and documentation) become constraints
for user interface design. UI generation can be trained or prompted to
satisfy these constraints, and downstream testing can verify them, much
as UICoder uses compiler and vision-language feedback in a self-training
loop \cite{wu2024}.

\textbf{Affordances and signifiers by construction:}
Norman\textquotesingle s distinction between perceived affordances and
conventions \cite{norman2013a} prescribes a critical
design principle: do not generate a control whose visual form does not
signal its action. A button must look pressable; a draggable handle must
look draggable; an editable field must signal editability. UI generation
can enforce these conventions at the component-library level, validating
the design before it reaches the user.

\textbf{Cognitive load and progressive disclosure:} Cognitive load
theory \cite{sweller1988} motivates a UI generation
policy of minimum cognitive footprint. The user interface exposes the capability the
first time the user\textquotesingle s dialogue articulates the
functional need. Progressive disclosure shifts from a design-time
concern to an artifact of user interaction and UI construction.

\textbf{Direct manipulation and the conversation--manipulation
tradeoff:} Shneiderman\textquotesingle s three direct-manipulation
criteria \cite{bigben1983} become an affordance for
design. The user interface ensures visible objects of interest, rapid
reversible incremental actions, and replacement of command syntax with
object manipulation.

\textbf{Accessibility as a generator invariant:} Automation of
accessibility ensures WCAG conformance
\cite{w3c2024a}. UI generation can treat WCAG success
criteria as hard constraints: every component emitted must pass
automated checks for contrast, focus order, ARIA (accessible rich internet application) roles, alt text,
keyboard reachability, and screen reading. LLM-driven generation will
make accessibility affordable across arbitrary domains, aligned with
ability-based personalization.

\textbf{Learnability as scaffolded emergence:} As the UI generation
grows with the user\textquotesingle s articulation of need, the user
quickly learns what each component is for: because they asked for it.
This collapses the classical learnability problem. The interface
functions become learnable through its provenance: every affordance
carries a back-link to the dialogue that justified it, providing the
recognition-over-recall property \cite{nielsen1994} for
free.

\section{Skills as Software Engineering Tools for
HCI}

A \textbf{skill} is a procedural specification of a capability, packaged
in a directory containing a skill.md file and any supporting assets
(schemas, scripts, templates, example components). The skill.md is
human-readable Markdown with structured frontmatter declaring when the
skill applies (a triggering description), what the AI must do
(procedural steps and constraints), and what counts as success
(verification criteria). Skills sit alongside the other AI-enabled
software engineering tools (Figure~\ref{fig:my_diagram}, top center) the agent draws on inside the Space to Think. Skills are declarative, inspectable, and
editable: an HCI researcher can write or audit a skill without
retraining a model, and the agent will load it on demand whenever the
triggering description matches the current task. We propose the
integration of HCI design knowledge repackaged as a small, opinionated
library of skills the agent invokes during UI generation (Figure~\ref{fig:my_diagram},
panel 4):

\begin{itemize}
\item
  \textbf{heuristics-nielsen.skill.md} encodes Nielsen\textquotesingle s
  ten usability heuristics \cite{nielsen1994} as a
  checklist the generator must satisfy: every action produces visible
  feedback; vocabulary matches the user\textquotesingle s domain
  (extracted from the dialogue); every destructive action is reversible;
  primary actions are recognizable from location and label without prior
  exposure; layouts obey a consistent component library.
\item
  \textbf{affordances-norman.skill.md} encodes Norman\textquotesingle s
  prescription \cite{norman2013a} that perceived form
  must signal real action. A button must look pressable; a draggable
  handle must look draggable; an editable field must signal editability.
  Because the generator emits structured component code rather than raw
  pixels, signifier compliance is enforced at the component-library
  level: any control failing the check is rejected before it reaches the
  user.
\item
  \textbf{accessibility-wcag.skill.md} encodes WCAG success criteria
  \cite{w3c2024a} as hard constraints. Every emitted
  component must pass automated checks for contrast, focus order, ARIA
  roles, alt text, and keyboard reachability. Conformance becomes a
  generator invariant rather than a release-time audit.
\item
  \textbf{ability-based-personalization.skill.md} encodes per-user
  accommodations beyond WCAG conformance, in the tradition of SUPPLE
  \cite{krzy2010}. Based on a user profile that the
  user articulates, this skill modifies UI construction and the other
  skills application. Personalization may include larger targets, longer
  dwell, simpler gestures, amplified contrast, type scaling, simplified
  layouts, plain-language labels, reduced surface area, longer time
  budgets, and step-by-step disclosure. This is uniquely enabled by the
  on-demand paradigm: each user interface is generated for one user.
\item
  \textbf{cognitive-load-progressive-disclosure.skill.md} encodes a
  minimum sufficient surface policy. The user interface exposes only the affordances
  the dialogue has so far justified; new capability appears the first
  time the user\textquotesingle s articulated need entails it.
\item
  \textbf{mixed-initiative.skill.md} encodes Horvitz\textquotesingle s
  principles \cite{horvitz1999} for the agentic
  affordances in the user interface: every action that affects the world is
  instrumented with confirmation, reversibility, an audit trail, and a
  visible confidence signal.
\item
  \textbf{direct-manipulation.skill.md} encodes
  Shneiderman\textquotesingle s three criteria
  \cite{bigben1983}: visible objects of interest, rapid
  reversible incremental actions, and replacement of command syntax with
  object manipulation, applied to each UI component.
\end{itemize}

The skills library is not a fixed set; HCI researchers contribute new
skills as research advances; practitioners audit, fork, and extend them;
organizations write skills encoding domain-specific conventions and
obtain a generator that follows them. Each artifact of the HCI
subdiscipline (design pattern, accessibility lint rule, layout
heuristic, etc.) becomes a runnable deliverable rather than a paper
hoping to influence practitioners. Skills carry properties that hand
design lacks:

\begin{itemize}
\item
  \textbf{Uniform}. Every component in every generated UI is checked,
  not only those a designer happened to remember.
\item
  \textbf{Inspectable}. A user can ask why a control behaves the way it
  does and the agent can cite the skill it followed. The same back-link
  supports the "Why this exists" affordance in panel 5.
\item
  \textbf{Editable}. Updates ship at the speed of HCI research, not the
  speed of pretraining. A regulator, an accessibility advocate, or a
  peer reviewer can read the skill and judge whether it captures the
  standard.
\item
  \textbf{Verifiable}. The same Markdown that instructs the generator
  can be compiled into automated tests. Conformance becomes part of HCI
  rather than a subjective audit.
\end{itemize}

\section{Future Work and Impact}

HCI knowledge changes at the speed of HCI research; whereas updating
model weights take months. Skills can be authored, reviewed,
version-controlled, and deployed in hours. Skills provide auditability;
a \textbf{skill.md} is a reviewable text file. Organizations and
individuals can layer skills without forking the underlying model.
On-demand UI provides a cognitive extension of the Space to Think, and a
natural mechanism for turning HCI design knowledge into procedural
constraints on generation. Good HCI becomes a property of a process, not
of a product; the field\textquotesingle s craft becomes runnable,
inspectable, and shareable.

LLM-generated code is notoriously difficult to verify, and the same is
true of the user interfaces it produces. This is not a new defect
introduced by automated generation but a consequence of
HCI\textquotesingle s longstanding validation problem that even
human-designed interfaces resist proof of "good design". Current
practice relies on user studies, formative and summative evaluations,
and heuristic walkthroughs that are expensive, slow, and still
inconclusive on learnability, error resistance, and long-term usability.
Future work is needed to develop formal validation criteria, automated
heuristic and accessibility checks that are machine interpretable. The
development of predictive models for cognitive load and learnability
will enable the ability to automate the validation of generated user
interfaces. We anticipate that verification skills will emerge as a
knowledge base alongside generation skills, becoming a future executable
validation toolkit.

Our research team will extend this work over the next year by
open-sourcing our initial skills library and applying it within our
ongoing Space to Think research to enable on-demand UIs that surface,
explore, and externalize chain-of-thought reasoning in Human-AI
interaction. At a five-year horizon, we expect HCI to have a skills
ecosystem comparable to today\textquotesingle s component libraries:
open repositories, organizational forks, peer-reviewed contributions. A
new HCI paper will routinely ship a skill alongside its findings, and
reviewers will expect a runnable artifact. The unit of design-knowledge
transfer will shift from the heuristic checklist to the executable
skill. At a ten-year horizon, we anticipate that pre-designed UIs will
be rare for heterogeneous, idiosyncratic tasks. Accessibility will
become a default that every user receives as the UI generation leverages
an accessibility-wcag skill and validates conformance. At a
twenty-five-year horizon, the most consequential change will be
conceptual. Cognitive properties will be assembled, and design will be
the discipline of shaping that assembly. The skills library may become
the field\textquotesingle s accumulated craft: machine-readable,
executable, and open. We may look back on the era of the fixed
application the way we now look back on the era of the form letter.

\begin{acks}
Thanks to Na Meng for your review and support of this paper. Thanks to Dave Cadorette and Vince Traglia for your support over the last two years enabling the exploration of this topic. 
\end{acks}

\bibliographystyle{ACM-Reference-Format}
\bibliography{paper-references}

\appendix

\section{Figure 1 Detailed Description}
\label{sec:my_appendix}

\subsection{Overview}

The figure shows a cycle of 6 panels.

\begin{enumerate}
    \item Space to think $\rightarrow$ panel 2
    \item Shared task model $\rightarrow$ panels 1, 3
    \item Generation pipeline $\rightarrow$ panel 5
    \item Automatic HCI principles enforcement $\rightarrow$ panel 5
    \item Generated user interface $\rightarrow$ panel 6
    \item Bi-directional coupling $\rightarrow$ panel 1
\end{enumerate}

\noindent There are two additional boxes.

\begin{itemize}
    \item AI-enabled software engineering tools $\rightarrow$ panel 2
    \item Result (below the entire diagram)
\end{itemize}

\subsection{Panel 1. Space to Think}
\label{sec:panel1}

Natural language dialogue to articulate intent, goals, and preferences.
\begin{itemize}
    \item User: I want to track my ETF holdings and see how my active picks are doing.
    \item Response: So you'd like a watchlist for long-term ETCs and performance tracking for active picks. Anything else?
    \item User: yes, compare to benchmarks and alert me when SPY drops 3\% in a day.
    \item Response: Got it. I'll set up performance charts, comparisons, and an alert configuration
\end{itemize}

\vspace{\baselineskip}
\noindent Dialogue Decomposition (AI)
\begin{itemize}
    \item Goals \& Sub-Goals
    \item Constraints \& Preferences
    \item Domain Concepts
    \item Actions \& Information Needs
    \item Success Criteria
\end{itemize}

\vspace{\baselineskip}
\noindent Panel 1 $\rightarrow$ ``\nameref{sec:panel2}''.

\subsection{AI-Enabled Software Engineering Tools}
\label{sec:aitools}

\begin{itemize}
    \item LLM Reasoning \& Planning
    \item Task \& Data Modeling
    \item UI Generation (Components + Code)
    \item Tooling, Libraries, \& Frameworks
    \item Testing \& Verification
\end{itemize}

\vspace{\baselineskip}
\noindent AI-Enabled Software Engineering Tools $\rightarrow$ ``\nameref{sec:panel2}''.

\subsection{Panel 2. Shared Task Model}
\label{sec:panel2}

Single source of truth.

Structured representation of the user's task and dialogue state.
\begin{itemize}
    \item Data \& Entities (e.g., positions, prices, alerts)
    \item Views \& Components (e.g., table, chart, config panel)
    \item Interactions \& Actions (e.g., filter, compare, alert)
    \item Provenance \& History (link to dialogue turns)
\end{itemize}

\vspace{\baselineskip}
\noindent Panel 2 $\rightarrow$ ``\nameref{sec:panel3}''.

\subsection{Panel 3. Generation Pipeline}
\label{sec:panel3}

\begin{itemize}
    \item Interpret Intent from Dialogue
    \item Plan UI \& Information Architecture
    \item Select Components \& Interactions
    \item Generate Code (HTML/React/...) \& Styles
    \item Instantiate UI
\end{itemize}

\vspace{\baselineskip}
\noindent Panel 3 $\rightarrow$ ``\nameref{sec:panel5}''.

\subsection{Panel 4. Automatic HCI Principles Enforcement}
\label{sec:panel4}

Procedural application of design knowledge.

\begin{itemize}
    \item Visibility of system status $\rightarrow$ Show feedback, data freshness
    \item Match to real world $\rightarrow$ Use domain language \& concepts
    \item User control \& freedom $\rightarrow$ Undo, edit, remove, rollback
    \item Consistency \& standards $\rightarrow$ Component library \& patterns
    \item Error prevention $\rightarrow$ Confirm destructive actions
    \item Recognition rather than recall $\rightarrow$ Labels, icons, defaults, tooltips
    \item Flexibility \& efficiency $\rightarrow$ Filters, shortcuts, multi-views
    \item Aesthetic \& minimalist design $\rightarrow$ Progressive disclosure, declutter
    \item Error recovery $\rightarrow$ Clear recovery \& help actions
    \item Help \& documentation $\rightarrow$ Contextual help \& provenance
    \item Accessibility (WCAG) $\rightarrow$ Contrast, focus order, ARIA, alt text,
keyboard navigation
\end{itemize}

\vspace{\baselineskip}
\noindent Automated checks \& AI critic verify compliance.

\par
\noindent Panel 4 $\rightarrow$ ``\nameref{sec:panel5}''.

\newpage
\subsection{Panel 5. Generated User Interface}
\label{sec:panel5}

Extension of thinking.

This panel has a dashboard with four boxes of information.

\begin{itemize}
    \item ETF Watchlist: a table with four columns (ETF, Price, 1D\%, YTD \%) and rows for four different stock symbols (VOO, VTI, VXUS, AGG). Last updated at 9:30 AM.
    \item Active Picks vs Benchmark: a line graph with two lines, Portfolio and SPY (Benchmark). The lines start at 0\% in January and rise through June, with a small dip in April. The Portfolio rises faster, ending at +10\% compared to Benchmark which ends at +5\%. Top contributors are NVDA +2.31\% and AVGO +1.28\%.
    \item Alerts: SPY Price Drop (alert me when SPY drops 3\% in a day) toggle switch is ON. Button ``Edit Alert''.
    \item Why this exists: you asked to be alerted when SPY drops 3\% in a day. Button ``View in conversation''.
\end{itemize}

\subsection{Panel 6. Bi-Directional Coupling}
\label{sec:panel6}

Think $\rightarrow$ UI $\rightarrow$ Think

\begin{itemize}
    \item \textbf{Refine in Conversation:} Ask follow-up questions, change coals, add constraints, request new views.
    \item \textbf{Interact Directly:} Filter, sort, edit, annotate, compare, drill down, configure alerts.
    \item \textbf{Updates Task Model:} UI Interactions update the shared model and inform next steps in dialogue.
\end{itemize}
    
\subsection{Result}
\label{sec:result}

A living, malleable interface.

\begin{itemize}
    \item The UI grows with your thinking
    \item Preserves provenance and spatial memory
    \item Good HCI by construction, not by chance
\end{itemize}

\end{document}